

\input harvmac.tex
\def\sdtimes{\mathbin{\hbox
{\hskip2pt\vrule height 4.1pt depth -.3pt width .25pt
\hskip-2pt$\times$}}}
\def\w{\omega_2}
\def\tws{\tilde\omega^*_2}
\def\l{\lambda}
\def\pl{\lambda^\prime}
\def\tl{\tilde\lambda}
\def\tpl{\tilde\lambda^\prime}
\def\bz{\bar z}
\long\def\optional#1{}
\lref\edmin{E. Witten, ``On the Landau-Ginzburg Description of N=2
Minimal Models'', {\it Int. J. Mod. Phys.} {\bf A9} (1994) 4783.}
\lref\moore{G. Moore, ``Atkin-Lehner Symmetry'', {\it Nucl. Phys.}
{\bf B293} (1987) 139.}
\lref\dixon{L. Dixon, ``Some World-Sheet Properties of Superstring
Compactifications, On Orbifolds and Otherwise'', Lectures given at
the 1987 ICTP Summer Workshop in High Energy Physics and Cosmology.}
\lref\spectrum{S. Kachru and E. Witten, ``Computing the Complete Massless
Spectrum of a Landau-Ginzburg Orbifold'', {\it Nucl. Phys.}
{\bf B407} (1993) 637, hepth/ 9307038.}
\lref\dkfirst{J. Distler and S. Kachru, ``(0,2) Landau-Ginzburg Theory'',
{\it Nucl. Phys.} {\bf B413} (1994) 213, hepth/9309110.}
\lref\dksecond{J. Distler and S. Kachru, ``Singlet Couplings and (0,2)
Models '', {\it Nucl. Phys.} {\bf B430} (1994) 13, hepth/9406090.}
\lref\nextus{E. Silverstein and E. Witten, forthcoming.}
\lref\gepner{D. Gepner, ``Spacetime Supersymmetry in Compactified String
Theory and Superconformal Models'', {\it Nucl. Phys.}
{\bf B296} (1986) 493; ``Exactly Solvable String Compactifications
on Manifolds of SU(N) Holonomy'', {\it Phys. Lett.} {\bf B199} (1987)
380.}
\lref\zamfat{A.B. Zamalodchikov and V.A. Fateev,``Nonlocal (parafermion)
currents in two-dimensional conformal quantum field theory and
self-dual critical points in $Z_N$-symmetric statistical systems'',
{\it Sov. Phys. JETP} {\bf 62} (2) (1985) 215.}
\lref\gsw{M. Green, J. Schwarz, and E. Witten, {\it Superstring Theory}
v.2, Cambridge University Press, Cambridge (1987) p. 441.}
\lref\zamfatII{A.B. Zamalodchikov and V.A. Fateev, ``Disorder Fields
in Two-Dimensional Conformal Quantum Field Theory and N=2 Extended
Supersymmetry'', {\it Sov. Phys.} JETP {\bf 63} (5) (1986) 913.}
\lref\distgreene{J. Distler and B. Greene,``Some Exact Results on the
Superpotential from Calabi-Yau Compactifications'', {it Nucl. Phys.}
{\bf B309} (1988) 295.}
\lref\bpz{A. Belavin, A. Polyakov, and A.B. Zamolodchikov, ``Infinite
Conformal Symmetry in Two-Dimensional Quantum Field Theory'', {\it Nucl. Phys.}
{\bf B241} (1984) 333.}
\lref\gradryz{I.S. Gradshteyn and I. M. Ryzhik, {\it Table of Integrals,
Series, and Products}, 5th ed. (Academic Press, Boston, 1994).}
\lref\dotpott{V. Dotsenko,``Critical Behavior and Associated Conformal
Algebra of the $Z_3$ Potts Model'',{\it Nucl. Phys.}{\bf B235}(1984) 54 }
\lref\phases{E. Witten, ``Phases of N=2 Theories in Two Dimensions'',
{\it Nucl. Phys.} {\bf B403} (1993) 159.}
%
\Title{\vbox{\hbox{PUPT--1519}\hbox{\tt hep-th/9503150}}}
{\vbox{\centerline{Miracle at the Gepner Point}
	\vskip2pt\centerline{}}}
\centerline{Eva Silverstein\footnote{$^\dagger$}
{silver@puhep1.princeton.edu}}
\bigskip\centerline{Joseph Henry Laboratories}
\centerline{Jadwin Hall}
\centerline{Princeton University}\centerline{Princeton, NJ 08544 USA}
\vskip .3in

A four-point function of $E_6$ singlets, of interest in elucidating
the moduli space of (0,2) deformations of the quintic string vacuum,
is computed using analytic and numerical methods.
The conformal field theory amplitude satisfies the requisite selection
rules and monodromy conditions, but the integrated string amplitude
vanishes. Together with selection rules coming from the spacetime
R-symmetry \dksecond~, this demonstrates the flatness of the
gauge-singlet spacetime superpotential
through fourth order.  Relevance to the more general program of determining
the (0,2) moduli space and superpotential is discussed.

\Date{03/95} 

\newsec{Introduction and Motivation}
The quintic hypersurface in ${\bf CP}^4$ is one of the best-studied
string vacua.  In the large radius (field theory) limit, it is
given by a nonlinear sigma model describing strings propagating
on a Calabi-Yau manifold $K$ constructed as follows (see \gsw~ for a review).
There are
five complex coordinates $s^i$ subject to one scaling relation
\eqn\scaling{s^i\sim \lambda s^i}
for any complex number $\lambda$.  The hypersurface is given by
the vanishing of a homogeneous degree five polynomial in the $s^i$:
\eqn\hyp{G(s^i)=G_{ijklm}s^is^js^ks^ls^m=0.}

The parameters $r$ and $G_{ijklm}$ determining the size and complex
structure of the manifold are true moduli of the (2,2) string vacuum
\dixon.  In addition to varying the complex structure of the manifold,
one can deform the tangent bundle $T$ to produce a more general rank 3
vector bundle $V$ over $K$ as follows.  A tangent vector $t\in T$
to the quintic is given by five complex numbers $t^i$ subject to
the equivalence $t^i\cong t^i+\lambda s^i$ and the constraint
$t^i{\partial G\over{\partial s^i}}=0$.  A vector $v\in V$ will be
given by five complex numbers $v^i$ subject to the equivalence
$v^i\cong t^i+\lambda s^i$ and the constraint
$v^i\bigl({\partial G\over{\partial s^i}}+H_i\bigr)=0$, for five
quartic polynomials
$H_i=H_{i,j_1j_2j_3j_4}s^{j_1}\dots s^{j_4}$ satisfying $s^iH_i=0$.
After subtracting the 25 linear redefinitions of the $s^i$
one finds 101 parameters in $G_{ijklm}$ and 224 in $H_{i,j_1j_2j_3j_4}$.

The quintic can be realized by a linear sigma model which contains
the Kahler parameter $r$ and the complex structure and bundle parameters
$G_{ijklm}$ and $H_{i,j_1j_2j_3j_4}$ as coupling constants \phases.
The deformations given by $H_{i,j_1j_2j_3j_4}$ break $(2,2)$ worldsheet
supersymmetry down to $(0,2)$.
For $r>>0$ one recovers the nonlinear sigma model on the quintic
hypersurface in the infrared.  For $r<<0$ one obtains a Landau-Ginsburg
orbifold.
For a special choice of the defining polynomial $G$, namely
\eqn\geppoly{G_{Gepner}(s^j)=\sum_i {(s^i)^5\over 5},}
there is substantial evidence that
the Landau-Ginsburg orbifold is equivalent to a tensor product
of $N=2$ minimal models, otherwise known as a Gepner model \gepner~
(see \edmin~ and references therein).

For each parameter described above one finds a space-time chiral
superfield in the massless spectrum of the model.
In particular, at large radius, the string vacuum
arising from the quintic
hypersurface in ${\bf CP}^4$ has 326 $E_6$ singlets, which
decompose into states arising from 1 Kahler modulus,
101 complex structure deformations, and 224 deformations of the tangent
bundle.  At very small radius, one finds 301 singlets in the untwisted sector
of the Landau-Ginsburg orbifold and 25 singlets with the same quantum
numbers as the ``missing" singlets in a twisted sector \spectrum~ \dkfirst.
The trace part of this the $5\times 5$ matrix of twisted singlets corresponds
to the Kahler modulus, while the traceless part gives 24 ``twisted singlets''.
At infinite radius all 325 singlets
are moduli: the space-time superpotential has 325 flat directions
corresponding to the complex structure and bundle deformations.

It was observed in \dksecond~ that a term in the spacetime superpotential
quartic in the twisted singlets would satisfy the spacetime R-symmetry
selection rule.  Using a global $SU(5)\times SU(5)$ symmetry
under which the worldsheet coupling constants as well as fields transform,
the contribution can be restricted to one independent amplitude \dksecond.
At the Gepner point, this satisfies the N=2 minimal
model selection rules, suggesting a possible obstruction to deforming in
the 24 twisted singlet
directions at small radius.  In this note we explicitly compute
this amplitude, relevant to a more general study of the (0,2) landscape,
and discover that it vanishes upon integration
over the locations of the vertex operators on the worldsheet
(to a convincing numerical precision).  Thus at least through fourth
order we can deform in all 224 extra singlet directions at LG after
all, in
keeping with more general arguments for flatness of (0,2) moduli
forthcoming in \nextus.

We will begin by setting up the computation in section 2, reviewing
how Gepner model amplitudes break up into free boson correlation functions
and parafermion amplitudes.  For the quintic, the parafermion
theory is (a sector of) the c=4/5 N=0 minimal model, and in section
3 we solve the differential equations satisfied by that part of the
amplitude and
find the unique nonzero linear combination of solutions which have trivial
monodromies.  In section 4 we assemble the pieces and report on the
numerical evaluation of the amplitude.

\newsec{Structure of the amplitude}

Gepner models are obtained by combining enough N=2 minimal
models (each with c=3k/(k+2))
to produce a c=9 internal conformal field theory which
is modular invariant and spacetime supersymmetric when combined
with the four spacetime dimensions \gepner.
The Gepner model for the quintic consists of 5 copies of the
c=9/5~ N=2 minimal model.  The N=2 minimal models each break up into
a free boson (which is the bosonization of the U(1) part of the N=2
superconformal algebra) and a c=4/5 parafermion theory \zamfatII.
The parafermion theory is a sector of the c=4/5 N=0 minimal model \zamfat.
We will recall the precise formulas for these relationships as we need them.

We are interested in a four-point function of ``twisted singlets" $S^i_j$.
In the Landau-Ginsburg orbifold the worldsheet fields include
complex bosons $s^i$ (which at large radius become the homogeneous
coordinates on ${\bf CP}^4$ described in the introduction) and
fermionic partners $\psi^i_\pm$, $\bar\psi^i_\pm$ \phases.  The model has
10 sectors, labeled by an integer $k=0,\dots,9$.  In this description
the twisted singlets are given by states of the form
\eqn\lgst{A_{\bar{i}j}\bar\psi^i_{-{1\over 5}}s^j_{-{3\over{10}}}|0>}
for constant matrix $A_{\bar{i} j}$, where $|0>$ is the
vacuum of the $k=3$ sector and the
subscripts indicate the mode of the worldsheet field used to create
the state \spectrum.  In the Gepner model, there is a discrete
symmetry group,
\eqn\gepsymm{(S_5\sdtimes Z_5^5)/Z_5}
coming from the fact that the
defining polynomial takes the simple form \geppoly.  There is
a $Z_5$ acting on each multiplet $(s^i,\psi^i_\pm)$
modulo a common $Z_5$ phase.  From the form of the states \lgst,
we see that under the Gepner model discrete symmetry group they
transform as
\eqn\geptrans{S^i_j\rightarrow e^{-{2 \pi i k_j \over 5}}
	e^{2 \pi i k_i \over 5}S^i_j}
where $e^{2 \pi i k_l \over 5}$ is an element of the lth $Z_5$.

The four-point function of interest is the following:
\eqn\fourpoint{\eqalign{
A=\Bigl<\epsilon^{\alpha\beta}
(V_1^5)^{-{1\over 2}}_{\alpha F}(z_1,\bz_1)
(V_2^5)^{-{1\over 2}}_{\beta F}(z_2,\bz_2)
(V_3^5)^{0}_{B}(z_3,\bz_3)
(V_4^5)^{-1}_{B}(z_4,\bz_4)
\Bigr>}}
The vertex operators in the Gepner model are determined
their dimensions, $U(1)$ charges, and
transformation properties under \gepsymm.
$$(V_1^5)^{-{1\over 2}}_{\alpha F}(z_1,\bz_1)=
S_\alpha(\bz_1) e^{-{\phi(\bz_1)\over 2}}
\Phi^0_{-2,-2;1,1}\otimes \Phi^1_{-1,0;2,1}\otimes \Phi^1_{-1,0;2,1}
\otimes\Phi^1_{-1,0;2,1}\otimes\Phi^2_{0,0;3,1}(z_1,\bz_2){\rm ,} $$
$$(V_2^5)^{-{1\over 2}}_{\beta F}(z_2,\bz_2)=
S_\beta(\bz_2) e^{-{\phi(\bz_2)\over 2}}
\Phi^1_{-1,0;2,1}\otimes\Phi^0_{-2,-2;1,1}\otimes\Phi^1_{-1,0;2,1}
\otimes\Phi^1_{-1,0;2,1}\otimes\Phi^2_{0,0;3,1}(z_1,\bz_1){\rm ,} $$
$$(V_3^5)^{0}_{B}(z_3,\bz_3)=
\Phi^1_{-1,0;1,0}\otimes \Phi^1_{-1,0;1,0}\otimes \Phi^0_{-2,-2;0,0}
\otimes\Phi^1_{-1,0;1,0}\otimes\Phi^2_{0,0;2,2}(z_3,\bz_3){\rm ,~ and} $$
$$(V_4^5)^{-1}_{B}(z_4,\bz_4)=
e^{-\phi(\bz_4)}
\Phi^1_{-1,0;1,0}\otimes\Phi^1_{-1,0;1,0}\otimes\Phi^1_{-1,0;1,0}
\otimes\Phi^0_{-2,-2;0,0}\otimes\Phi^2_{0,0;2,0}(z_4,\bz_4) $$
Here $S_\alpha$ is the spacetime spin field and $\phi$ is the standard
bosonization of the ghost number current.
The N=2 minimal model primary fields
 $\Phi^l_{q,s;\bar q,\bar s}$
(in the notation of \distgreene) break up into parafermion primary
fields $\phi^l_{q-s,{\bar q}-{\bar s}}$ times exponentials of the free
U(1) boson H, $e^{\alpha_{q,s} H} e^{\alpha_{\bar q,\bar s} \bar H}$,
where $\alpha_{q,s}={1\over \sqrt{15}}(-q+{5\over 2}s)$;
see \distgreene~ for a review.  The parafermion primary field
$\phi^l_{k,\bar k}$ has dimension
\eqn\dimpar{\eqalign{
&h^l_{k,\bar k}={{l(l+2)}\over{20}}-{k^2\over{12}};
\cr
&{\bar h^l_{k,\bar k}}={{l(l+2)}\over{20}}-{\bar k^2\over{12}}.}}
Under the Gepner model discrete symmetry group \gepsymm~ the
$N=2$ minimal model primary fields transform as
\eqn\primtransf{\Phi^l_{q,s;\bar q,\bar s}\rightarrow
e^{-i\pi(q+\bar q)/5}\Phi^l_{q,s;\bar q,\bar s}.}
{}From this formula it is easy to check that the above
vertex operators have the correct transformation properties
\geptrans~ under \gepsymm.

Note that the external and superconformal ghost pieces are independent
of $(z_3,\bar z_3)$.  Their correlation functions yield:
\eqn\ext{A_{ext}=\epsilon^{\alpha\beta}
	\Bigl<S_\alpha(\bz_1)S_\beta(\bz_2)\Bigr>=
	{1\over(\bz_1-\bz_2)^{1\over 2}}}
\eqn\ghost{A_{ghost}=\Bigl<e^{-{\bar \phi \over 2}}(\bz_1)
	e^{-{\bar \phi \over 2}}(\bz_2)e^{-{\bar \phi}}(\bz_4)
	\Bigr>={1\over \bz^{1\over 4}_{12}}
	{1\over \bz^{1\over 2}_{24}}{1\over \bz^{1\over 2}_{14}}}
We use SL(2,C) to set $z_1$,$z_2$,
and $z_4$ to 0,1, and $\infty$,
remembering to include the Jacobian $|z_{12}|^2|z_{14}|^2|z_{24}|^2$
which is also independent of $z_3$.
Fixing  $z_1$,$z_2$, and $z_4$ this way the above contributions become
constants.  From the decomposition described above we find that
the parts of the free boson amplitude that depend on
$(z_3,\bar z_3)\equiv (z,\bar z)$ are:
\eqn\boson{A_H=|z_{13}|^{-{8\over 15}}|z_{23}|^{-{8\over 15}}
=|z|^{-{8\over 15}}|1-z|^{-{8\over 15}}.}

The parafermion amplitudes for the first four minimal model factors
are:
\eqn\paraA{P_1=\Bigl<\phi^0_{0,0}(z_1,\bz_1)\phi^1_{-1,1}(z_2,\bz_2)
			\phi^1_{-1,1}(z_3,\bz_3)\phi^1_{-1,1}(z_4,\bz_4)
			\Bigr>}
\eqn\paraB{P_2=\Bigl<\phi^1_{-1,1}(z_1,\bz_1)\phi^0_{0,0}(z_2,\bz_2)
			\phi^1_{-1,1}(z_3,\bz_3)\phi^1_{-1,1}(z_4,\bz_4)
			\Bigr>}
\eqn\paraC{P_3=\Bigl<\phi^1_{-1,1}(z_1,\bz_1)\phi^1_{-1,1}(z_2,\bz_2)
			\phi^0_{0,0}(z_3,\bz_3)\phi^1_{-1,1}(z_4,\bz_4)
			\Bigr>}
\eqn\paraD{P_4=\Bigl<\phi^1_{-1,1}(z_1,\bz_1)\phi^1_{-1,1}(z_2,\bz_2)
			\phi^1_{-1,1}(z_3,\bz_3)\phi^0_{0,0}(z_4,\bz_4)
			\Bigr>}
Because of the presence of the field $\phi^0_{0,0}$, the parafermion
identity field, each of these correlation functions reduces to
a three-point function equivalent by duality to those worked out
in \zamfat:
\eqn\exampleparathree{P_1=C_{1,1}|z_{23}|^{-{2\over 15}}
			|z_{34}|^{-{2\over 15}}|z_{24}|^{-{2\over 15}}}
where
\eqn\threept{C^2_{1,1}={\Gamma(1/5)\Gamma^3(3/5)
	            \over{\Gamma(4/5)\Gamma^3(2/5)}}}
and similarly for $P_2$, $P_3$, and $P_4$.
All together, the parafermion three-point functions contribute a factor
\eqn\totparathree{|z|^{-{4\over 15}}|1-z|^{-{4\over 15}}}
to the z,$\bz$-dependent part of the amplitude.
What remains is to compute the nontrivial fifth parafermion amplitude,
which will be the subject of the next section.

\newsec{The parafermion amplitude}
The nontrivial parafermion factor in the amplitude is
$$\bigl<\phi^2_{0,2}(0) \phi^2_{0,2}(1) \phi^2_{0,0}(z,\bz)
	\phi^2_{0,2}(\infty) \bigr>$$
Under the $Z_3 \times Z_3$ symmetry of the parafermion model,
$\phi^l_{k,\bar k} \rightarrow e^{{2 \pi i n_1 \over 3}{({k-\bar k\over 2})}}
e^{{2 \pi i n_2 \over 3}{({k+\bar k\over 2})}}\phi^l_{k,\bar k}$
\zamfat, so this amplitude satisfies
the selection rules.

The parafermion theory is a unitary theory with $c=4/5<1$, and so
must be a subset of one of the minimal models.  The $N=0$ series
of minimal models has $c=1-{6\over{m(m+1)}},\,\,m=3,4,\dots $.
The primary fields $\phi_{p,q}$ are
organized in the Kac table with dimensions
\eqn\dimmin{h_{p,q}={{(p(m+1)-qm)^2 -1} \over{4m(m+1)}}.}
In our case, $c=4/5\Rightarrow m=5$.
The parafermion primary fields, whose dimensions were given
in \dimpar, form a subset of the minimal model primary fields \dimmin~
for $m=5$.
In particular, our amplitude is equivalent to the N=0 c=4/5 minimal
model amplitude
$$\Bigl<\phi_{2,1}(0),\phi_{2,1}(1),\phi_{2,1}(z)
,\phi_{2,1}(\infty)\Bigr>
\Bigl<\phi_{3,3}(0),\phi_{3,3}(1),\phi_{2,1}(\bar z)
,\phi_{3,3}(\infty)\Bigr>$$
The minimal model primary field $\phi_{p,q}$ has a null state at
level $pq$ in its Verma module, and therefore satisfies a
differential equation of order $pq$.
Our four-point amplitude satisfies second-order
differential equations in z and $\bar z$ by virtue of the presence
of the $\phi_{2,1}$ field for each chirality \bpz:

\eqn\gendiff{\Bigl[{3\over{2 (2 \delta + 1)}}{d^2 \over{dz^2}} +
\sum_{i \ne 3}{{1 \over (z-z_i)}{d \over {dz}} - {\Delta_i \over{(z-z_i)^2}}}
+ \sum_{j<i}{{\delta + \Delta_{i,j}} \over {(z-z_i)(z-z_j)}}\Bigr]
G(z|z_1,z_2,z_4) = 0.}
where $\Delta_{1,2}=\Delta_1+\Delta_2-\Delta_4$ and $\delta={\rm dim}\phi_{2,1}
=2/5$, and where z is
replaced by $\bar z$ for the right-movers.  We set $z_1=0$, $z_2=1$,
and $z_4=\infty$ and substitute in the dimensions for the fields
given in \dimmin~ or equivalently \dimpar.  Then \gendiff~
can be rewritten as

\eqn\gendiff{z(1-z)G'' + {6\over 5}(1-2z)G' +
\Bigl[{12\over 5} \Delta-{6\over 5}({2\over 5}+\Delta)
-{6\over 5}\Delta({1\over z}+{1\over {1-z}})
\Big]G = 0}
where $\Delta$ is 2/5 for the holomorphic and 1/15 for the antiholomorphic
factors in the amplitude (for which z is also replaced by $\bz$).
The holomorphic factor has the two independent
solutions
\eqn\uone{U_1 = z^{3\over 5}(1-z)^{3\over 5}{_2F_1}(6/5,13/5,12/5;z)}
\eqn\utwo{U_2 = z^{-{4\over 5}}(1-z)^{-{4\over 5}}{_2F_1}(-8/5,-1/5,-2/5;z)}
and the antiholomorphic factor has the two independent solutions
\eqn\tuone{\tilde U_1 = \bz^{1\over 5}
		(1-\bz)^{1\over 5}{_2F_1}(4/5,7/5,8/5;\bz)}
\eqn\tutwo{\tilde U_2 = \bz^{-{2\over 5}}
                (1-\bz)^{-{2\over 5}}{_2F_1}(-2/5,1/5,2/5;\bz)}
where $_2F_1(a,b,c;z)$ is the standard hypergeometric function.

To obtain the physical correlation function, we must put the two
chiralities together in such a way as to leave the amplitude well-defined
on the complex plane, i.e. with trivial monodromies.  The amplitude
will be proportional to
\eqn\together{\left(\matrix{\tilde U_1&\tilde U_2\cr}\right)A
\left(\matrix{U_1\cr
	      U_2\cr}\right)}
where A is a $2\times 2$ constant matrix.
The hypergeometric functions in
\uone-\tutwo~ are given by convergent power series near $z=0$, so
we can read off the monodromies about z=0 from the prefactors:

\eqn\monzero{\eqalign{\left(\matrix{U_1\cr
              U_2\cr}\right) \rightarrow
\left(\matrix{e^{{3\over 5}2 \pi i}&0\cr
		0&e^{-{4\over 5}2 \pi i}\cr}\right)
\left(\matrix{U_1\cr
              U_2\cr}\right)}}
and
\eqn\tmonzero{\left(\matrix{\tilde U_1&\tilde U_2\cr}\right)\rightarrow
\left(\matrix{\tilde U_1&\tilde U_2\cr}\right)
\left(\matrix{e^{-{1\over 5}2 \pi i}&0\cr
                0&e^{{2\over 5}2 \pi i}\cr}\right)}
Requiring \together~ to remain invariant we learn that $a_{11}=a_{22}=0$.

To obtain the tranformations under monodromies about $z=\bz=1$
one makes use of the standard formula \gradryz
\eqn\hyptransf{\eqalign{_2F_1(a,b,c;z) =
&(1-z)^{-a}
{\Gamma(c)\Gamma(c-b-a)\over{\Gamma(c-a)\Gamma(c-b)}}
{_2F_1}(a,b,a+b-c+1;1-z)\cr
&+(1-z)^{c-a-b}{\Gamma(c)\Gamma(a+b-c)\over{\Gamma(a)\Gamma(b)}}
{_2F_1}(c-a,c-b,c-a-b+1;1-z)\cr}}
In \dotpott~ Dotsenko used \hyptransf~ to write down the
monodromy matrices for general hypergeometric solutions.  Using
his formulas we find the following monodromies
about $z=\bz=1$:
\eqn\monone{\eqalign{\left(\matrix{U_1\cr
              U_2\cr}\right) \rightarrow {e^{2 \pi i ({3\over 5})}
	\over {\lambda^\prime - \lambda}}
\left(\matrix{\pl-\l\w&\w -1\cr
	      \l\pl(1-\w)&\pl\w-\l\cr}\right)
\left(\matrix{U_1\cr
              U_2\cr}\right)}}

\eqn\tmonone{\left(\matrix{\tilde U_1&\tilde U_2\cr}\right)\rightarrow
\left(\matrix{\tilde U_1&\tilde U_2\cr}\right)
{e^{{{-2 \pi i} \over 5}}
	\over {\tilde\lambda^\prime - \tilde\lambda}}
\left(\matrix{\tpl-\tl\tws&\tl\tpl(1-\tws)\cr
              \tws-1&\tpl\tws-\tl\cr}\right)}
where $\w=e^{-2\pi i({2\over 5})}$,
$\l={\Gamma(-{2\over 5})\Gamma({6\over 5})\over
{\Gamma(-{8\over 5})\Gamma({12\over 5})}}$,
and
$\pl={\Gamma(-{2\over 5})\Gamma({13\over 5})\over
{\Gamma(-{1\over 5})\Gamma({12\over 5})}}$; and where
$\tws=e^{2\pi i({3\over 5})}$,
$\tl={\Gamma({2\over 5})\Gamma({4\over 5})\over
{\Gamma(-{2\over 5})\Gamma({8\over 5})}}$,
and
$\tpl={\Gamma({2\over 5})\Gamma({7\over 5})\over
{\Gamma({1\over 5})\Gamma({8\over 5})}}$.

One finds that
\eqn\onetransf{\eqalign{\left(\matrix{\tilde U_1&\tilde U_2\cr}\right)
	\left(\matrix{0&a_{12}\cr
	              a_{21}&0\cr}\right)
	  \left(\matrix{U_1\cr
              U_2\cr}\right)=
	\left(\matrix{\tilde U_1&\tilde U_2\cr}\right)
	\tilde M_1
	\left(\matrix{0&a_{12}\cr
	              a_{21}&0\cr}\right)
	M_1
	\left(\matrix{U_1\cr
              U_2\cr}\right)}}
(where $\tilde M_1$ and $M_1$ are the matrices that appear in \tmonone~ and
\monone) if and only if
\eqn\c{c\equiv{a_{21}\over a_{12}}=
	-{{(\tpl-\tl\tws)(\l\tl(1-\w))}\over
	  {(\pl-\l\w)(\tl\tpl(1-\tws))}}}
That is, up to scale, $\tilde U_1 U_2 + c \tilde U_2 U_1$ is the
unique solution satisfying the differential equations and monodromy
conditions.
\newsec{The integral}

Assembling the pieces, our amplitude is
\eqn\ampfull{\int d^2z |z|^{-{4\over 5}} |1-z|^{-{4\over 5}}
	(\tilde U_1 U_2 + c \tilde U_2 U_1)}
Note that because of the chiral nature of the internal amplitude
the integrand is not positive definite.  We computed this using
Mathematica's numerical integration routine {\it NIntegrate},
producing a vanishing result good to seven digits. In evaluating \ampfull~
we excised small regions containing the points 0, 1, and $\infty$
and evaluated their contribution separately.

This result supports the general arguments for flatness of
space-time superpotentials for $(0,2)$
linear sigma models to be explained in \nextus.
However, it would be more satisfying to obtain a direct analytical
understanding of the vanishing of this integrated amplitude.
Perhaps this could be attained by considering the integral
as an inner product between the left and right moving solutions,
and trying to understand why the two are orthogonal
analytically\foot{This hope is somewhat reminiscent of the argument,
involving Atkin-Lehner symmetry, for the vanishing
of the one-loop cosmological constant in certain non-space-time
supersymmetric vacua \moore;
I thank J. Distler for pointing this out to me.}.

\vskip 1in
\noindent
{\bf Acknowledgements}
I would like to thank J. Distler, S. Kachru, N. Nekrassov, and E. Witten
for discussions, and the NSF and AT$\&$T GRPW for support.
\listrefs

\end